\documentclass[11pt]{article}
\usepackage[margin=0.78in]{geometry}
\usepackage[T1]{fontenc}
\usepackage{mathptmx}
\usepackage{courier}
\usepackage{graphicx}
\usepackage{amsmath}
\usepackage{booktabs}
\usepackage{array}
\usepackage{xcolor}
\usepackage{hyperref}
\usepackage{tikz}
\usetikzlibrary{arrows.meta, positioning, fit, calc}
\usepackage{enumitem}
\usepackage{caption}
\usepackage{subcaption}
\usepackage{colortbl}
\usepackage{amssymb}
\usepackage[most]{tcolorbox}
\definecolor{figok}{HTML}{1D8A3A}
\definecolor{figbad}{HTML}{C0392B}
\newtcolorbox{samplebox}[2]{enhanced, colback=white, colframe=black, boxrule=0.6pt, arc=2pt,
  left=5pt, right=5pt, top=2pt, bottom=3pt, before skip=3pt, after skip=3pt, fontupper=\scriptsize,
  before upper={\noindent{\footnotesize\textbf{#1}\hfill #2}\par\vspace{-1pt}\noindent\rule{\linewidth}{0.5pt}\par\vspace{1pt}}}
\newcommand{\slab}[1]{\textbf{#1:}\ }
\newcommand{\ssep}{\par\vspace{-2pt}{\color{figedge}\rule{\linewidth}{0.4pt}}\par\vspace{-1pt}}
\newcommand{\tagok}[1]{{\sffamily\scriptsize\bfseries\color{figok}[#1]}}
\newcommand{\tagbad}[1]{{\sffamily\scriptsize\bfseries\color{figbad}[#1]}}
\newcommand{\tagmid}[1]{{\sffamily\scriptsize\bfseries\color{figbrown}[#1]}}
\newcommand{\cut}{{\color{figgray}[\ldots]}}
\usepackage{tabularx}
\newcolumntype{Y}{>{\centering\arraybackslash}X}
\newcolumntype{R}{>{\raggedleft\arraybackslash}X}
\newcolumntype{L}{>{\raggedright\arraybackslash}X}
\usepackage{multirow}
\usepackage{bm}
\usepackage[scaled=0.88]{helvet}
\definecolor{fignavy}{HTML}{1F3A5F}
\definecolor{figamber}{HTML}{E8A33D}
\definecolor{figamberlt}{HTML}{FDF3E1}
\definecolor{figbrown}{HTML}{7A5200}
\definecolor{figpanel}{HTML}{F7F8FA}
\definecolor{figedge}{HTML}{B8BCC4}
\definecolor{figslate}{HTML}{E6EAF0}
\definecolor{figgray}{HTML}{5A606B}
\tikzset{
  panel/.style={draw=figedge, fill=figpanel, rounded corners=4pt, line width=0.6pt},
  ptitle/.style={font=\sffamily\bfseries\footnotesize, text=fignavy, anchor=north west},
  card/.style={draw=figedge, fill=white, rounded corners=2.5pt, line width=0.5pt, font=\sffamily\scriptsize, align=center, inner sep=3pt},
  model/.style={fill=fignavy, draw=fignavy, rounded corners=3pt, text=white, font=\sffamily\bfseries\scriptsize, align=center, inner sep=4pt},
  hot/.style={draw=figamber, fill=figamberlt, rounded corners=2.5pt, line width=0.6pt, font=\sffamily\scriptsize, align=center, inner sep=3pt},
  grp/.style={draw=figamber, dashed, rounded corners=4pt, line width=0.7pt},
  note/.style={font=\sffamily\itshape\tiny, text=figgray, align=center},
  flow/.style={-{Latex[length=1.6mm]}, line width=0.6pt, draw=figgray},
}
\usepackage{url}

\hypersetup{
  colorlinks=true,
  linkcolor=black,
  citecolor=black,
  urlcolor=blue,
  pdftitle={Dual-Node NVIDIA DGX Spark over Tailscale: A Remote-Access Testbed for Distributed LLM Training and Cyber-Threat-Intelligence Fine-Tuning}
}

\makeatletter\g@addto@macro{\UrlBreaks}{\do0\do1\do2\do3\do4\do5\do6\do7\do8\do9\do a\do b\do c\do d\do e\do f}\makeatother
\title{\textbf{Dual-Node NVIDIA DGX Spark over Tailscale: A Remote-Access\\
Testbed for Distributed LLM Training and\\
Cyber-Threat-Intelligence Fine-Tuning}}

\author{
Vasanth Iyer\\
Department of Computer Science and Digital Technologies\\
Grambling State University\\
\texttt{iyerv@gram.edu}
}
\date{August 2026; revised October 2026 (v2)}

\begin{document}
\maketitle

\begin{abstract}
Compact AI systems make local language-model experimentation increasingly accessible, yet practical evidence for multi-node training on desktop-class accelerators remains limited. This report presents a proof-of-concept deployment of distributed NanoChat pretraining across two NVIDIA DGX Spark systems, each with a GB10 Grace Blackwell system-on-chip and 128~GB of unified memory, administered remotely over a Tailscale mesh VPN and connected for training by a dedicated 200~Gb/s QSFP56 direct fiber link. PyTorch \texttt{torchrun}, DDP, and NCCL were configured with one process per node, a depth-20 NanoChat model, a device batch of 32 sequences per node, a 2{,}048-token context, and eight gradient-accumulation steps, giving 1{,}048{,}576 tokens per optimizer step for the 896.5-million-parameter model. The run took 4{,}980 steps over about four days, averaging 63.4~s of measured training time per step (about 16{,}540 tokens/s) and processing about 5.22 billion tokens. We document link configuration, container setup, interface binding, a step-zero evaluation bug that triggered NCCL timeouts, checkpointing, and troubleshooting lessons, as a reproducibility reference for small labs.

We also built a cybersecurity fine-tuning dataset from 77 CISA advisories (338 training, 37 validation conversations) and compared a baseline SFT checkpoint against a CTI-augmented checkpoint on 17 held-out prompts with an Ollama-hosted LLM judge: CTI-specific categories improved while general-knowledge categories regressed, for a small overall change from 2.06 to 2.29 on a 0--10 scale. Version~2 adds a stage-2 fine-tune with a MITRE ATT\&CK technique dictionary under a deterministic evaluation: in-dictionary ID-to-name lookup reaches 98.5\%, yet none of 50 held-out IDs is named correctly, knowledge of held-out threat groups, software, campaigns, and vulnerabilities stays near zero, and general ability drops measurably. The same cluster supports a 400-level AI course (CS~426) and a query engine for CompTIA Security+ POGIL activities in CBS~255, showing modest local infrastructure can serve both research and teaching. The study establishes feasibility rather than a scaling-efficiency claim. Runbook and scripts are available (see Code Availability).

\noindent\fbox{\parbox{\dimexpr\linewidth-2\fboxsep-2\fboxrule}{\small
\textbf{Changes in v2.} An audit of v1 against the run's checkpoint metadata, training reports, shell histories, and saved evaluation files found the following errors, corrected here. (1)~v1 omitted gradient accumulation: the optimizer step was 1{,}048{,}576 tokens, not 131{,}072, so total tokens were about 5.22~billion (not 653~million) and throughput about 16{,}540 tokens/s (not 1{,}890); the 3.5$\times$ single-node comparison is withdrawn (Section~\ref{sec:training}). (2)~The LLM judge was \texttt{llama3:latest}, not \texttt{llama3.1:8b} (Section~\ref{sec:evalpipe}). (3)~Each SFT run was 485 steps on one GPU at 1{,}048{,}576 tokens/step, not about 2{,}700 steps at 8{,}192 tokens/step, and the CTI checkpoint was a one-stage SFT from the base model with the CISA rows forming about 0.03\% of the mixture (Section~\ref{sec:ctistructure}). (4)~Pretraining ran with \texttt{save\_every=-1}, so only the final checkpoint exists. (5)~Previously unreported: sampling settings, 15 of 34 judged answers truncated at a timeout, and an SFT learning-rate schedule bug. The v1 judge scores themselves reproduce from the saved files. New in v2: the parameter count (896{,}533{,}746), the nanochat commit and container digest, and a stage-2 ATT\&CK-dictionary fine-tuning study with a deterministic evaluation (Section~\ref{sec:stage2}).}}

\end{abstract}

\section{Introduction}

Training and adapting language models locally is attractive for universities, small laboratories, and security-sensitive projects that cannot depend exclusively on cloud services. Local infrastructure provides control over data movement, repeatability, software configuration, and operational cost. However, most published distributed-training examples assume datacenter clusters with mature schedulers, shared storage, and high-end multi-GPU servers. There is comparatively little practical documentation showing how two compact desktop AI systems can be turned into a reliable distributed training environment, or how such a system can be reused beyond a single research workload -- let alone made conveniently and securely reachable by students and administrators who are rarely on the same physical network as the hardware.

This report studies a two-node NVIDIA DGX Spark deployment used to pretrain a NanoChat transformer, fine-tune it for cyber-threat-intelligence (CTI) tasks, and support two university courses, all administered remotely over a Tailscale mesh VPN. The work is intentionally framed as a systems proof of concept rather than a new learning algorithm. Its central question is:

\begin{quote}
\textit{Can two DGX Spark systems, linked by a dedicated NVIDIA fiber connection for training traffic and a Tailscale mesh VPN for remote administration, be configured as a practical, reproducible distributed LLM training testbed, and can the resulting infrastructure support a complete domain-specific fine-tuning and evaluation workflow alongside routine instructional use?}
\end{quote}

The contributions are:
\begin{enumerate}[topsep=2pt, itemsep=1pt]
  \item a documented two-node DGX Spark architecture with separate Tailscale-managed remote-access and NCCL data planes;
  \item reproducible container, network, NCCL, and \texttt{torchrun} procedures for depth-20 NanoChat training;
  \item an operational analysis of startup timeouts, persistent addressing, long-running sessions, checkpointing, and rank monitoring;
  \item a downstream CTI case study including CISA-to-SFT conversion, automated response generation, paired checkpoint evaluation, and LLM-judge aggregation;
  \item (new in v2) a stage-2 fine-tuning study that teaches the model MITRE ATT\&CK technique identifiers, evaluated deterministically on in-dictionary and held-out IDs, held-out CTI entities, and held-out general prompts;
  \item an educational-use case study in which the same cluster supports a 400-level AI course and a certification-aligned POGIL/PBQ query engine; and
  \item an explicit accounting of what is measured, estimated, and not resolved, including (in v2) an audit of the v1 claims against the run's artifacts, so the report's claims are not overstated.
\end{enumerate}

\section{Background}

\subsection{NanoChat}
\label{sec:nanochat-bg}
NanoChat is a compact full-stack language-model training repository that includes tokenization, base pretraining, supervised fine-tuning, evaluation, inference, and an interactive chat interface~\cite{nanochat}. Its depth parameter controls the transformer family used by the training scripts. The experiment documented here used \texttt{--depth=20} on both distributed ranks. v1 reported only the configured depth; for v2 the parameter count was computed from the tensors of the saved base and SFT checkpoints, which agree with one another, with nanochat's own parameter counter, and with its training report (Table~\ref{tab:params}). Almost half of the parameters are in nanochat's ten value-embedding tables. The forward pass costs about $2.89\times10^{9}$ FLOPs per token. The $\approx$31-million-parameter model mentioned in Section~\ref{sec:embedding} is a different, smaller configuration.

\begin{table}[h]
\centering
\small
\begin{tabularx}{0.72\linewidth}{Lrr}
\toprule
\textbf{Group} & \textbf{Parameters} & \textbf{Share} \\
\midrule
Value embeddings (10 tables, $32{,}768\times1{,}280$) & 419{,}430{,}400 & 46.8\% \\
Transformer blocks (20) & 393{,}217{,}200 & 43.9\% \\
Token embedding & 41{,}943{,}040 & 4.7\% \\
Output head & 41{,}943{,}040 & 4.7\% \\
\textcolor{figgray}{Scalar gates} & \textcolor{figgray}{66} & \textcolor{figgray}{--} \\
\midrule
\rowcolor{figamberlt}\textbf{Total} & \textbf{896{,}533{,}746} & \textbf{100\%} \\
\bottomrule
\end{tabularx}
\caption{\textbf{Parameter count of the depth-20 model} (new in v2), computed from the tensors of the saved base and SFT checkpoints, which are identical in size.}
\label{tab:params}
\end{table}

\subsection{PyTorch DDP and NCCL}
PyTorch Distributed Data Parallel (DDP) replicates a model across processes, computes gradients on local batches, and synchronizes gradients before each optimizer update~\cite{ddp}. \texttt{torchrun} launches the distributed processes across nodes~\cite{torchrun}, while NCCL provides GPU-oriented collective communication~\cite{nccl}. In this deployment, one process was launched per node and NCCL communication was explicitly bound to the dedicated fiber interface using \texttt{NCCL\_SOCKET\_IFNAME}.

\subsection{Related Work}
Distributed data-parallel training of large models is well studied on datacenter-class clusters. Horovod introduced ring-allreduce-based data parallelism for commodity GPU clusters~\cite{horovod}; ZeRO~\cite{zero} and PyTorch's Fully Sharded Data Parallel (FSDP)~\cite{fsdp} extend data parallelism with memory-efficient parameter, gradient, and optimizer-state sharding for models beyond single-device capacity. These systems primarily target multi-GPU servers or datacenter interconnects (InfiniBand, NVLink) rather than the point-to-point, dual-node topology of a small laboratory deployment. Compact, fully open training stacks such as nanoGPT~\cite{nanogpt} and NanoChat~\cite{nanochat} make model internals -- tokenizer, pretraining loop, SFT, inference -- directly inspectable, which is valuable both for reproducible systems research and for classroom use (Section~\ref{sec:education}). On the desktop accelerator side, NVIDIA's DGX Spark line packages a Grace Blackwell superchip with unified memory in a workstation form factor~\cite{dgxspark}; to our knowledge this report is among the first to document a dual-node DGX Spark deployment for distributed pretraining -- made remotely accessible over a Tailscale mesh VPN -- as opposed to single-node inference or fine-tuning. For downstream domain adaptation, our own related work explores instruction-tuned, retrieval- and graph-augmented fine-tuning of domain-specific LLMs for MITRE ATT\&CK-aligned evaluation~\cite{cyberllm}; the present report differs in scope, treating the CISA fine-tuning pipeline as a downstream validation of the distributed-training deployment itself rather than as an attempt to optimize CTI task accuracy, and it reports the resulting checkpoint's weak factual grounding candidly rather than only its favorable deltas. Evaluation follows the LLM-as-judge methodology of Zheng et al.~\cite{llmjudge}. On the pedagogical side, the POGIL activities used in Section~\ref{sec:education} follow the process-oriented guided-inquiry framework of Moog and Spencer~\cite{pogil}.

\section{Physical and Network Architecture}
\label{sec:architecture}

\subsection{Laboratory Deployment}
\begin{figure}[h]
\centering
\includegraphics[width=0.62\textwidth]{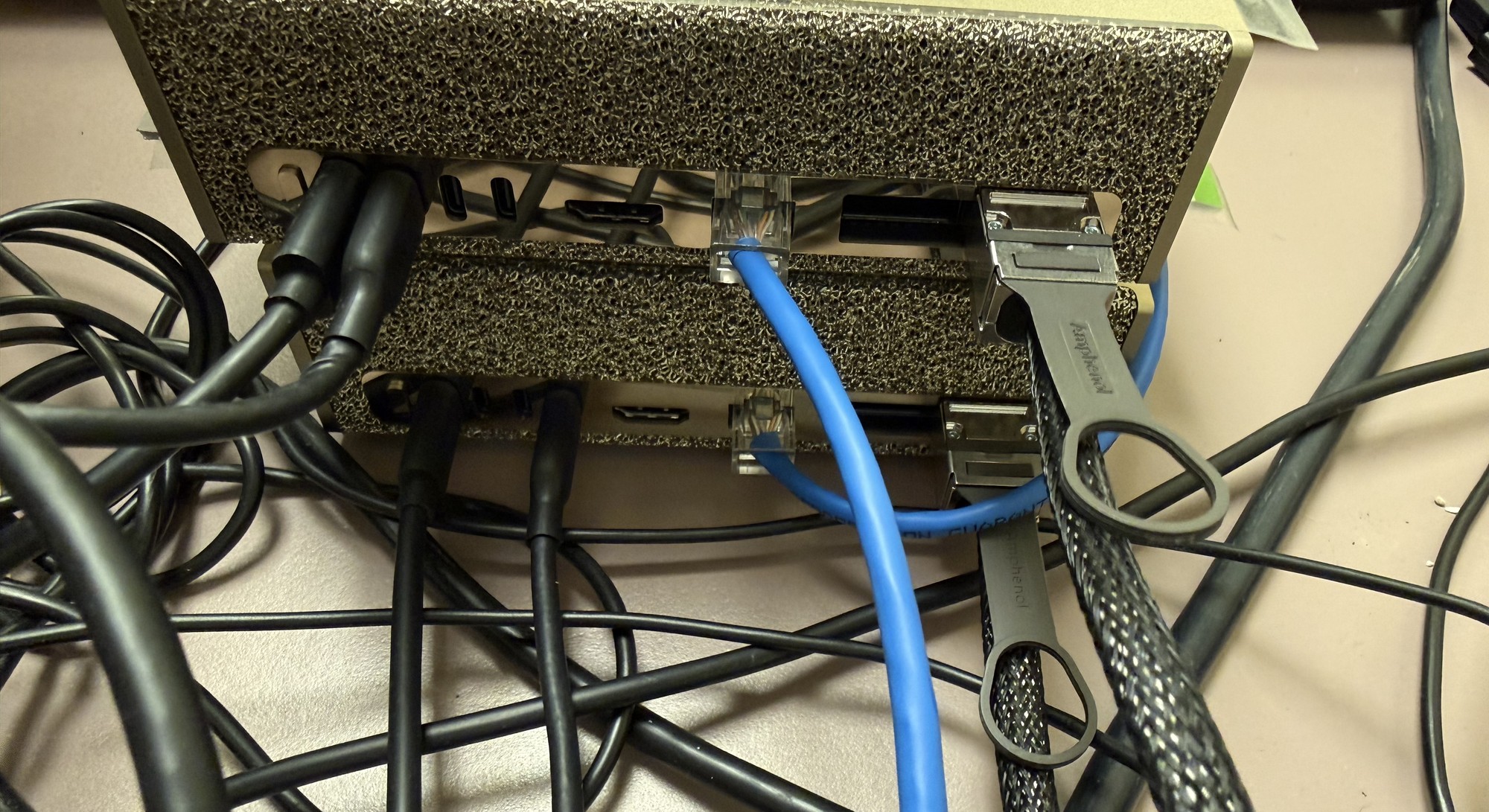}
\caption{\textbf{Laboratory deployment.} Rear view of the two stacked NVIDIA DGX Spark systems used in the proof-of-concept deployment, showing the point-to-point QSFP cable that joins their ConnectX ports for the training plane (right) and the Ethernet connections (blue).}
\label{fig:lab}
\end{figure}

Figure~\ref{fig:lab} shows the two DGX Spark systems installed in the Grambling State University laboratory. The photograph is included as deployment evidence rather than a performance measurement. The systems were stacked temporarily during setup; the operational configuration should maintain manufacturer-recommended clearance and airflow.

\subsection{Management Plane and Training Plane}
Remote administration used a Tailscale mesh VPN (with SSH over the tailnet), while gradient synchronization was routed over a separate point-to-point network. The runbook assigns \texttt{spark-9fcd} as rank~0 and \texttt{spark-4654} as rank~1. A dedicated QSFP56 fiber link connected interface \texttt{enp1s0f1np1} on both systems using the private, non-routable subnet \texttt{192.168.100.0/24}. Jumbo frames were configured with MTU~9000. Keeping the Tailscale-managed remote-access plane and the NCCL training plane on physically distinct interfaces -- rather than layering training traffic over the VPN, or remote access over the training fiber -- is the central architectural decision this report documents; Figure~\ref{fig:planes} summarizes the separation.

\begin{figure}[h]
\centering
\begin{tikzpicture}[x=1cm,y=1cm]
  \node[panel, minimum width=12.6cm, minimum height=5.1cm, anchor=north west] at (0,0) {};
  \node[ptitle] at (0.15,-0.12) {Management plane and training plane};
  \node[card, fill=figslate, minimum width=2.6cm] (ws) at (2.0,-1.15) {Administrator\\workstation};
  \node[card, fill=figslate, minimum width=2.6cm] (st) at (10.6,-1.15) {Student devices\\(CS 426, CBS 255)};
  \node[card, minimum width=4.1cm, minimum height=0.55cm] (ts) at (6.3,-1.15) {\textbf{Tailscale mesh VPN}\\SSH over the tailnet};
  \draw[flow, dashed] (ws) -- (ts);
  \draw[flow, dashed] (st) -- (ts);
  \node[model, minimum width=3.0cm, minimum height=1.05cm] (a) at (2.4,-3.55) {spark-9fcd\\[-1pt]{\sffamily\mdseries\tiny rank 0 / master \textperiodcentered{} 192.168.100.11}};
  \node[model, minimum width=3.0cm, minimum height=1.05cm] (b) at (10.2,-3.55) {spark-4654\\[-1pt]{\sffamily\mdseries\tiny rank 1 / worker \textperiodcentered{} 192.168.100.10}};
  \draw[flow, dashed] (ts.south) -- ++(0,-0.45) -| (a.north);
  \draw[flow, dashed] (ts.south) -- ++(0,-0.45) -| (b.north);
  \node[note, anchor=west] at (6.42,-1.66) {management traffic only};
  \draw[figamber, line width=2.2pt] (a.east) -- (b.west);
  \node[hot, anchor=south] at (6.3,-3.43) {200 Gb/s QSFP56 fiber \textperiodcentered{} {\ttfamily\tiny enp1s0f1np1}};
  \node[note] at (6.3,-3.8) {NCCL / DDP gradients \textperiodcentered{} MTU 9000};
  \node[grp, fit=(a)(b), inner sep=7pt] (g) {};
  \node[font=\sffamily\bfseries\tiny, text=figbrown, anchor=north east] at ($(g.south east)+(-0.1,-0.03)$) {training plane};
  \node[note, anchor=south west] at (0.25,-4.95) {Ollama judge and the CBS 255 quiz model also run on rank 0, reached over the tailnet};
\end{tikzpicture}
\caption{\textbf{Separate management and training planes.} Remote administration and classroom traffic reach both nodes over a Tailscale mesh VPN (dashed), while gradient synchronization runs only over the dedicated point-to-point fiber link (amber).}
\label{fig:planes}
\end{figure}

\begin{table}[h]
\centering
\small
\begin{tabularx}{\linewidth}{llL}
\toprule
& \textbf{Component} & \textbf{Configuration} \\
\midrule
\multirow{4}{*}{Hardware} & Nodes & \texttt{spark-9fcd} (rank 0) and \texttt{spark-4654} (rank 1) \\
& Accelerator & One NVIDIA GB10 Grace Blackwell SoC per node \\
& Unified memory & 128 GB per node \\
& Training link & 200 Gb/s QSFP56 direct fiber \\
\midrule
\multirow{4}{*}{Network} & Training interface & \texttt{enp1s0f1np1} \\
& Training subnet & 192.168.100.10/24 and 192.168.100.11/24 \\
& MTU & 9000 \\
& Remote access & Tailscale mesh VPN (single tailnet) \\
\midrule
\multirow{3}{*}{Software} & Container & NVIDIA NGC PyTorch 25.10-py3 (torch 2.9.0a0, Python 3.12.3) \\
& NanoChat & commit \texttt{dc54a1a} plus two local patches (Section~\ref{sec:repro}) \\
& Launcher & PyTorch \texttt{torchrun}, two nodes, one process per node \\
\bottomrule
\end{tabularx}
\caption{\textbf{Two-node system configuration} documented by the deployment artifacts.}
\label{tab:config}
\end{table}

\section{Reproducible Deployment Procedure}

\subsection{Persistent Link Configuration}
The two hosts did not necessarily use the same Linux network manager. The runbook therefore checks both \texttt{nmcli} and \texttt{netplan}, then creates a persistent profile with static addresses and MTU~9000. Link validation includes bidirectional ping, \texttt{ethtool} verification of 200{,}000~Mb/s full duplex, and a nonfragmenting 8{,}972-byte ping payload.

\subsection{Container Configuration}
Each node launched a host-networked NVIDIA PyTorch container. The host network is required so the container can see the dedicated fiber interface directly. Persistent mounts preserve the NanoChat cache, tokenizer, data, checkpoints, and repository across disposable containers.

\begin{quote}
\ttfamily\small
\noindent
docker run --gpus all -it --rm \textbackslash\\
\phantom{xx}--ulimit memlock=-1 --ulimit stack=67108864 \textbackslash\\
\phantom{xx}--network host \textbackslash\\
\phantom{xx}-v /home/nsf-iomt-llm/.cache/nanochat:/root/.cache/nanochat \textbackslash\\
\phantom{xx}-v \textasciitilde/token-training/nanochat:/workspace/nanochat \textbackslash\\
\phantom{xx}nvcr.io/nvidia/pytorch:25.10-py3 bash
\end{quote}
\begin{center}\small Listing~1: Container launch used on each node.\end{center}

The runbook additionally installs Python dependencies and builds NanoChat's Rust tokenizer. Dataset and tokenizer preparation is performed on both nodes unless their caches are explicitly synchronized.

\subsection{Step-Zero Evaluation Correction}
A critical operational failure occurred before the first training step. The original base-training condition invoked a slow bits-per-byte evaluation at step zero:

\begin{quote}\ttfamily\small step \% args.eval\_every == 0\end{quote}

The deployment changed it to:

\begin{quote}\ttfamily\small step \% args.eval\_every == 0 and step > 0\end{quote}

Without this guard, the first multi-node attempt waited approximately 17 minutes and then failed through the NCCL watchdog before training began. This local modification is a patch against the upstream NanoChat repository and must be reapplied when starting from a fresh clone.

\subsection{Distributed Launch}
Both nodes exported the same communication settings:

\begin{quote}\ttfamily\small
export NCCL\_SOCKET\_IFNAME=enp1s0f1np1\\
export NCCL\_TIMEOUT=3600\\
export TORCH\_NCCL\_BLOCKING\_WAIT=1\\
export WANDB\_MODE=disabled
\end{quote}

Rank~0 launched with:

\begin{quote}\ttfamily\small
torchrun --nproc\_per\_node=1 --nnodes=2 --node\_rank=0 \textbackslash\\
\phantom{xx}--master\_addr=192.168.100.11 --master\_port=29501 \textbackslash\\
\phantom{xx}-m scripts.base\_train -- \textbackslash\\
\phantom{xx}--depth=20 --device-batch-size=32 \textbackslash\\
\phantom{xx}--run=\textless RUN\_NAME\textgreater
\end{quote}

Here \texttt{<RUN\_NAME>} denotes an operator-assigned identifier used for checkpoint directory naming, not a placeholder for missing information. Rank~1 used the same command with \texttt{--node\_rank=1}. v1 reconstructed this command from the rank-0 log and listed \texttt{--save-every=500}; the configuration recorded in the final checkpoint's metadata shows \texttt{save\_every=-1}, so the run saved no intermediate checkpoints and only the step-4{,}980 checkpoint exists. The total batch size was left at nanochat's default of 1{,}048{,}576 tokens, which the launch command does not show (Section~\ref{sec:training}). Both sessions ran inside \texttt{tmux} to survive SSH interruption. Only rank~0 emitted step/loss/throughput output; rank~1 activity was confirmed through \texttt{nvidia-smi} and process memory allocation.

\section{Training Experiment}
\label{sec:training}

\subsection{Configuration}
The production experiment used a depth-20 model, a 2{,}048-token context, and a device batch size of 32 on each node. NanoChat fixes the optimizer batch through its total-batch-size setting (default 1{,}048{,}576 tokens) and makes up the difference with gradient accumulation. With two ranks, one forward/backward micro-step covers $2 \times 32 \times 2{,}048 = 131{,}072$ tokens, so each optimizer step accumulated eight micro-steps:

\begin{equation}
N_{\text{tokens/step}} = 2\ \text{nodes} \times 32\ \text{sequences/node} \times 2048\ \text{tokens/sequence} \times 8\ \text{accumulation steps} = 1{,}048{,}576.
\label{eq:tokens}
\end{equation}

v1 reported 131{,}072 tokens per step, which is one micro-step; the checkpoint metadata records \texttt{total\_batch\_size} 1{,}048{,}576 and \texttt{device\_batch\_size} 32. Over 4{,}980 steps, the run processed $4{,}980 \times 1{,}048{,}576 = 5{,}221{,}908{,}480$ tokens, matching the token count in nanochat's training report.

\subsection{Measured Throughput}
The training loop's own timer records 315{,}750.5~s of training time over 4{,}980 steps, or 63.4~s per step. v1's 69.4~s per step is wall-clock time (about four days divided by 4{,}980 steps) and includes the periodic evaluations. Aggregate token throughput over training time is therefore:

\begin{equation}
\text{throughput} = \frac{1{,}048{,}576}{63.4} \approx 16{,}540\ \text{tokens/s}.
\label{eq:throughput}
\end{equation}

\noindent At the end of pretraining, the validation loss was 1.0713 bits per byte (minimum during training 0.9967), the CORE estimate was 0.0919, and peak memory was 71.8~GB.

\begin{table}[h]
\centering
\small
\begin{tabularx}{\linewidth}{llccRRcl}
\toprule
& \textbf{Run} & \textbf{GPUs} & \textbf{Device batch} & \textbf{Tokens/step} & \textbf{Tokens/s}$\uparrow$ & \textbf{Steps} & \textbf{Wall clock} \\
\midrule
\multirow{3}{*}{Pretrain} & \textcolor{figgray}{d12 validation$^\dagger$} & \textcolor{figgray}{1} & \textcolor{figgray}{32} & \textcolor{figgray}{65{,}536} & \textcolor{figgray}{36{,}700} & \textcolor{figgray}{2{,}520} & \textcolor{figgray}{about 10 h} \\
& \textcolor{figgray}{d20 single node$^\dagger$} & \textcolor{figgray}{1} & \textcolor{figgray}{32} & \textcolor{figgray}{65{,}536} & \textcolor{figgray}{540} & \textcolor{figgray}{9{,}960} & \textcolor{figgray}{about 14 d (est.)} \\
& \cellcolor{figamberlt}\textbf{d20 dual node} & \cellcolor{figamberlt}\textbf{2} & \cellcolor{figamberlt}\textbf{32} & \cellcolor{figamberlt}\textbf{1{,}048{,}576} & \cellcolor{figamberlt}\textbf{16{,}540} & \cellcolor{figamberlt}\textbf{4{,}980} & \cellcolor{figamberlt}\textbf{about 4 d (measured)} \\
\midrule
\multirow{2}{*}{SFT} & Baseline, d20 & 1 & 32 & 1{,}048{,}576 & -- & 485 & about 15 h \\
& CTI, d20 & 1 & 32 & 1{,}048{,}576 & -- & 485 & about 15 h \\
\bottomrule
\end{tabularx}
\caption{\textbf{Training phases,} corrected in v2 from the checkpoint metadata and training reports. The production run is highlighted. $^\dagger$: from the project evaluation document; not checked against logs, and may share v1's omission of gradient accumulation.}
\label{tab:phases}
\end{table}

v1 compared the four-day dual-node run with a 14-day single-node projection and described a 3.5$\times$ reduction. That comparison is withdrawn: the single-node figure was an estimate, no single-node log exists, and its tokens-per-step figure has the same form as the undercounted dual-node value. v1 also stated that adding a node doubled the global batch. Because nanochat holds the total batch fixed and varies gradient accumulation instead, this was an artifact of the undercount.

Both SFT runs used one GPU (DDP world size 1) with 16 accumulation steps and inherited the base run's 1{,}048{,}576-token batch, for 485 steps each (about 509 million tokens). v1's figures of about 2{,}700 steps at 8{,}192 tokens per step could not have covered the roughly 1.07-million-row SFT mixture even once. The roughly 15-hour duration v1 reported is consistent with the container lifetimes (at most 16.5~h and 17.1~h, including false starts).

\subsection{Controlled Benchmark for Future Work}
A publication-quality scaling claim requires matched experiments after warm-up, all at the same 1{,}048{,}576-token optimizer batch:
\begin{enumerate}[topsep=2pt, itemsep=1pt]
  \item one node, device batch 32, 16 accumulation steps;
  \item two nodes, device batch 32, 8 accumulation steps; and
  \item two nodes, device batch 16, 16 accumulation steps, to separate the effect of micro-batch size from that of node count.
\end{enumerate}
For each condition, such a study should report median and interquartile-range step time over at least 100--300 steady-state steps, aggregate and per-node tokens/s, GPU utilization, link traffic, memory use, and power draw. Scaling efficiency would then be reported as
\begin{equation}
E_2 = \frac{T_1/T_2}{2} \times 100\%,
\label{eq:efficiency}
\end{equation}
where $T_1$ and $T_2$ are matched one-node and two-node step times. This report does not include that benchmark; Table~\ref{tab:phases} instead documents the single production run actually performed, with estimated and measured values clearly distinguished.

\section{Cyber-Threat-Intelligence Fine-Tuning Case Study}

\subsection{CISA-to-SFT Dataset Construction}
The \texttt{cisa\_to\_sft.py} script converts a source CSV containing 77 CISA advisories~\cite{cisa}, cleaned text, and mapped MITRE ATT\&CK identifiers~\cite{mitre} into NanoChat-compatible JSONL conversations. It generates four forms of training interaction: advisory summaries, TTP enumeration, positive technique-presence questions, and negative technique-presence questions. Text is cleaned and truncated to fit the SFT context budget, ATT\&CK IDs are validated with a regular expression, and manually verified title overrides repair advisories whose titles are difficult to extract automatically.

\begin{figure}[t]
\centering
\begin{tikzpicture}[x=1cm,y=1cm]
  \node[panel, minimum width=6.15cm, minimum height=7.0cm, anchor=north west] at (0,0) {};
  \node[ptitle] at (0.12,-0.12) {(a) v1: One-stage CTI SFT};
  \node[card, minimum width=5.2cm] (a1) at (3.07,-0.95) {77 CISA advisories\\cleaned text + mapped ATT\&CK IDs};
  \node[card, minimum width=5.2cm] (a2) at (3.07,-1.95) {\texttt{cisa\_to\_sft.py}\\338 train + 37 validation conversations};
  \node[hot, minimum width=5.2cm] (a3) at (3.07,-2.95) {general SFT mixture ($\approx$1.07M rows)\\CISA rows $\approx$0.03\%, one epoch};
  \node[model, minimum width=5.2cm] (a4) at (3.07,-4.0) {SFT from base step 4{,}980\\{\sffamily\mdseries\tiny baseline and CTI runs, 485 steps each}};
  \node[card, minimum width=5.2cm] (a5) at (3.07,-5.05) {\texttt{batch\_eval.py}: 17 prompts\\sampled, $T{=}0.6$, 120 s timeout};
  \node[card, minimum width=5.2cm] (a6) at (3.07,-6.1) {\texttt{judge\_eval.py}, \texttt{llama3:latest}\\factual score + key facts};
  \foreach \i/\j in {a1/a2,a2/a3,a3/a4,a4/a5,a5/a6} \draw[flow] (\i) -- (\j);
  \begin{scope}[xshift=6.45cm]
  \node[panel, minimum width=6.15cm, minimum height=7.0cm, anchor=north west] at (0,0) {};
  \node[ptitle] at (0.12,-0.12) {(b) v2: Stage-2 dictionary fine-tune};
  \node[card, minimum width=2.45cm] (b1) at (1.55,-0.95) {ATT\&CK v20\\STIX 2.1};
  \node[card, minimum width=2.45cm] (b1b) at (4.6,-0.95) {338 CISA\\conversations};
  \node[hot, minimum width=2.45cm] (b2) at (1.55,-1.95) {dictionary\\5{,}598 rows};
  \node[note] at (1.55,-2.55) {84 IDs excluded};
  \node[card, minimum width=2.45cm] (b2b) at (4.6,-1.95) {+ SmolTalk\\replay};
  \node[model, minimum width=5.2cm] (b3) at (3.07,-3.2) {Stage-2 SFT from d20-original\\{\sffamily\mdseries\tiny 49--405 steps, two or three row orders per arm}};
  \node[grp, minimum width=5.45cm, minimum height=1.62cm] (g) at (3.07,-5.2) {};
  \node[font=\sffamily\bfseries\tiny, text=figbrown, anchor=north west] at ($(g.north west)+(0.08,-0.04)$) {eval v2, greedy};
  \node[card, minimum width=1.55cm] (e1) at (1.42,-5.3) {T\# lookup\\100 + 50};
  \node[card, minimum width=1.55cm] (e2) at (3.07,-5.3) {CTI\\118 + 37};
  \node[card, minimum width=1.55cm] (e3) at (4.72,-5.3) {general-40};
  \node[note] at (3.07,-6.45) {exact-name match, key facts, \texttt{llama3:latest} judge};
  \draw[flow] (b1) -- (b2); \draw[flow] (b2) -- (b3); \draw[flow] (b1b) -- (b2b); \draw[flow] (b2b) -- (b3);
  \draw[flow] (b3) -- (g.north);
  \end{scope}
\end{tikzpicture}
\caption{\textbf{CTI fine-tuning and evaluation workflows.} \textbf{(a)} The v1 case study: CISA conversations are mixed into the general SFT mixture and trained in one stage from the base checkpoint (corrected structure; Section~\ref{sec:ctistructure}), then scored by an LLM judge on 17 sampled prompts. \textbf{(b)} The v2 stage-2 study: an ATT\&CK technique dictionary and the CISA conversations are fine-tuned from the baseline checkpoint and evaluated deterministically on in-dictionary and held-out IDs, held-out and seen CTI items, and held-out general prompts (Section~\ref{sec:stage2}).}
\label{fig:workflow}
\end{figure}

With seed 42 and a 10\% validation split, the resulting artifacts contain 338 training and 37 validation conversations (375 total). The average combined user/assistant character count is approximately 897 for training and 793 for validation, computed directly from the JSONL files. The 37 validation conversations were not used by any training or judging run in v1. Nearly all of their advisories also appear in the training rows, so they measure recall of seen content rather than generalization (Section~\ref{sec:stage2} uses them that way).

\subsection{Structure of the CTI Fine-Tune (corrected in v2)}
\label{sec:ctistructure}
v1 did not state how the CTI checkpoint was trained relative to the baseline. The shell histories and checkpoint metadata show that both checkpoints are \emph{one-stage} SFT runs from the base checkpoint at step 4{,}980. The baseline (``d20-original'') used nanochat's general SFT mixture of about 1.07 million rows. The CTI checkpoint (``d20'') reran the same 485-step recipe with the 338 CISA conversations inserted into that mixture, where they made up about 0.03\% of the rows for one epoch. It is not a continuation of the baseline. The two runs reached essentially the same validation loss (0.40233 and 0.40231 bits per byte), so the contrast in Table~\ref{tab:results} is between two near-identical runs whose only difference is a very small CTI dose. Both runs also started with a fresh optimizer state: the base optimizer shard had been renamed before the successful baseline run, so its warm-start setting had no effect. Finally, nanochat's SFT learning-rate schedule at this commit does not clamp its progress variable, so on the last step the learning-rate multiplier can turn slightly negative (we observed $-0.02$ in a later 49-step run). Both reference runs used the unclamped schedule; their logs were not retained, so the size of any overshoot is unknown. A clamp was added to the course-kit copy during the stage-2 work.

\subsection{Held-Out Evaluation Set}
The separate held-out set contains 17 prompts that are not copied verbatim from the generated training conversations. Each item includes a category, natural-language prompt, ground-truth answer, and verifiable key-fact list. The distribution is shown in Table~\ref{tab:heldout}.

\begin{table}[h]
\centering
\small
\begin{tabularx}{0.5\linewidth}{lLc}
\toprule
& \textbf{Category} & \textbf{Items} \\
\midrule
\multirow{3}{*}{CTI} & Technique lookup & 5 \\
& Name to technique & 3 \\
& Threat actor & 4 \\
\midrule
\multirow{2}{*}{General} & General concept & 3 \\
& General-chat control & 2 \\
\midrule
& \textbf{Total} & \textbf{17} \\
\bottomrule
\end{tabularx}
\caption{\textbf{Held-out evaluation set} (v1) by category.}
\label{tab:heldout}
\end{table}

\subsection{Automated Evaluation Pipeline}
\label{sec:evalpipe}
The evaluation pipeline is fully scripted:
\begin{enumerate}[topsep=2pt, itemsep=1pt]
  \item \texttt{batch\_eval.py} launches NanoChat's interactive SFT CLI through \texttt{pexpect}, submits all prompts, captures responses, detects repetition-loop timeouts, resets conversation state, and saves results incrementally;
  \item the script is run once for d20-original and once for d20;
  \item \texttt{merge\_responses.py} pairs baseline and CTI-SFT outputs by item identifier; and
  \item \texttt{judge\_eval.py} sends each response, ground truth, and key-fact list to an Ollama-hosted \texttt{llama3:latest} judge~\cite{ollama} at temperature 0.1, then reports factual-accuracy scores and key-fact counts by category and overall, following the LLM-as-judge approach of Zheng et al.~\cite{llmjudge}.
\end{enumerate}

\noindent\textbf{Corrections in v2.} v1 named the judge as \texttt{llama3.1:8b}. The judge script that produced the saved results specifies \texttt{llama3:latest}, and re-judging the saved responses with the unchanged prompt reproduces 33 of the 34 recorded scores exactly with \texttt{llama3:latest} (mean absolute difference 0.03), against 23 of 34 with \texttt{llama3.1:8b} (0.65). The judge ran on the rank-0 node itself. v1 also did not report how answers were generated: the CLI sampled at temperature 0.6 with top-$k$ 50, at most 256 new tokens, and seed 42, one turn per prompt. Of the 34 saved answers, 15 (7 baseline, 8 CTI) hit the driver's 120-second timeout; they were saved with a truncation marker and judged as saved.

\subsection{Evaluation Results}
Table~\ref{tab:results} reproduces the aggregate results from the supplied evaluation report. All three CTI-specific categories improved, while general concepts and general-chat controls declined.

\begin{table}[h]
\centering
\small
\begin{tabularx}{0.78\linewidth}{llYYYY}
\toprule
& \textbf{Category} & $\boldsymbol{n}$ & \textbf{Baseline}$\uparrow$ & \textbf{CTI-SFT}$\uparrow$ & \textbf{Delta} \\
\midrule
\multirow{3}{*}{CTI} & Name to technique & 3 & 0.67 & \textbf{2.00} & +1.33 \\
& Technique lookup & 5 & 0.00 & \textbf{0.40} & +0.40 \\
& Threat actor & 4 & 0.50 & \textbf{1.50} & +1.00 \\
\midrule
\multirow{2}{*}{General} & General concept & 3 & \textbf{6.33} & 4.67 & $-$1.67 \\
& Control general chat & 2 & \textbf{6.00} & 5.50 & $-$0.50 \\
\midrule
\rowcolor{figamberlt} & \textbf{Overall} & \textbf{17} & 2.06 & \textbf{2.29} & \textbf{+0.24} \\
\bottomrule
\end{tabularx}
\caption{\textbf{LLM-judge factual accuracy} (0--10, higher is better) for the v1 case study. Delta is CTI-SFT minus baseline; bold marks the higher of the two scores. Answers were sampled and 15 of 34 were truncated (Section~\ref{sec:evalpipe}).}
\label{tab:results}
\end{table}

The results indicate a measurable domain-vocabulary shift but weak factual grounding. The CTI gains are small in absolute terms, and the largest regression occurs in general concepts. The overall difference of +0.24 is not statistically conclusive for 17 items and one judge model. The result is best interpreted as evidence that the downstream pipeline functioned end to end and that a small CTI dose measurably altered behavior, not as evidence of a strong domain model.

\textbf{Reproduction and noise (new in v2).} Re-judging the saved responses reproduces every number in Table~\ref{tab:results}. Regenerating the answers without the timeout gives 2.47 for the baseline and 2.65 for the CTI checkpoint. About 10 of 17 answers per model are word-for-word identical to the v1 run; the differences fall on the answers v1 had truncated. Sampled answers make single runs of this evaluation noisy: two stage-2 runs with the same data, the same 49 steps, and the same validation loss to four decimal places scored 3.06 and 1.77 overall, because sampling diverges once the weights differ even slightly, and individual 0--10 judge scores then swing by up to 9 points. We therefore treat a single run's 17-prompt score as uncertain by roughly $\pm1.3$, and its 2--5-prompt category scores as unable to rank models. Section~\ref{sec:stage2} switches to greedy decoding and larger test sets for this reason.

\subsection{Stage-2 Fine-Tuning with an ATT\&CK Technique Dictionary (new in v2)}
\label{sec:stage2}
The one-stage CTI checkpoint did not learn technique identifiers (technique lookup 0.40/10 in Table~\ref{tab:results}). We therefore ran a second, short SFT stage starting from the baseline checkpoint (d20-original, step 485), with device batch 16 and the learning rate at 0.2 of the SFT rate.

\textbf{Data.} The 338 CISA conversations were combined with a dictionary built from MITRE ATT\&CK Enterprise v20.0~\cite{mitre,strom2018attack} in STIX~2.1 format~\cite{mitrestix,stix21}: 622 technique IDs, three tasks each (ID$\to$name, name$\to$ID, ID$\to$tactic), and three phrasing templates, for 5{,}598 rows. Excluded from the dictionary were the IDs used by the 17-prompt evaluation together with their sub-techniques (34 IDs) and 50 further IDs held out for testing (36 sub-techniques and 14 top-level techniques). General SmolTalk conversations~\cite{smollm2} were replayed alongside, at 1.2278 replay rows per uploaded row unless stated. An automated audit confirmed that no held-out ID appears with its name anywhere in the fine-tuning data. Twenty of the 50 held-out IDs do occur as bare IDs, mostly inside lists of technique IDs in the CISA answers; the base model's pretraining data was not audited.

\textbf{Evaluation.} All answers are greedy, so generation is deterministic: re-evaluating a model reproduced every answer exactly (6~models $\times$ 322 answers), leaving the judge as the only source of run-to-run noise. Four test sets were used:
\begin{itemize}[topsep=2pt, itemsep=1pt]
  \item \emph{T\# lookup}: 100 in-dictionary and 50 held-out IDs in an evaluation-only phrasing, scored as correct if the answer contains the official technique name;
  \item \emph{held-out CTI}: 118 items about entities that appear in no training or evaluation data (40 ATT\&CK groups, 40 software entries, 18 campaigns, and 20 entries from CISA's Known Exploited Vulnerabilities catalog~\cite{kev});
  \item \emph{seen CTI}: 37 items asking, in new wording, about CISA advisories that are in the training data (20 technique questions, 17 summaries); and
  \item \emph{general-40}: 40 held-out general prompts across science, mathematics, everyday knowledge, world knowledge, language, and computing, none sharing an 8-word run with SmolTalk or any other evaluation set.
\end{itemize}
CTI and general answers are scored with a deterministic key-fact match over the first 150 words and with the unchanged \texttt{llama3:latest} judge prompt.

\begin{table}[h]
\centering
\small
\setlength{\tabcolsep}{4pt}
\begin{tabularx}{\linewidth}{llrYYYYYY}
\toprule
& & & \multicolumn{2}{c}{\textbf{General-40}} & & \multicolumn{2}{c}{\textbf{Seen CTI}} & \\
\cmidrule(lr){4-5}\cmidrule(lr){7-8}
& \textbf{Arm} & \textbf{Steps} & Facts$\uparrow$ & Judge$\uparrow$ & \textbf{T\# dict}$\uparrow$ & Facts$\uparrow$ & Judge$\uparrow$ & \textbf{CISA techn.}$\uparrow$ \\
\midrule
\textcolor{figgray}{--} & \textcolor{figgray}{d20-original (baseline)} & \textcolor{figgray}{485} & \textcolor{figgray}{37.9} & \textcolor{figgray}{5.42} & \textcolor{figgray}{0.0} & \textcolor{figgray}{0.0} & \textcolor{figgray}{0.43} & \textcolor{figgray}{0.0} \\
\midrule
\multicolumn{9}{l}{\textit{\textcolor{figgray}{Stage 2 from d20-original}}} \\
\midrule
CISA & CISA $\times$3 (replay 5) & 49 & \underline{35.8} & \underline{5.00} & 0.0 & \underline{27.0} & \underline{2.08} & \underline{50.0} \\
\midrule
\multirow{2}{*}{One file} & CISA + dict.\ $\times$3 & 206 & 32.5 & 4.53 & 82.0 & 15.3 & 1.96 & 26.7 \\
& CISA + dict.\ $\times$5 & 226 & 31.0 & 4.76 & 98.5 & 22.1 & 1.64 & 39.2 \\
\midrule
\multirow{3}{*}{Two files} & Dict.\ $\times$5, CISA $\times$3 & 221 & 30.6 & 4.57 & \underline{99.0} & 21.6 & 1.72 & 36.7 \\
& \cellcolor{figamberlt}\textbf{Dict.\ $\times$5, CISA $\times$5} & \cellcolor{figamberlt}226 & \cellcolor{figamberlt}\textbf{37.7} & \cellcolor{figamberlt}\textbf{5.15} & \cellcolor{figamberlt}97.0 & \cellcolor{figamberlt}\textbf{35.1} & \cellcolor{figamberlt}\textbf{2.85} & \cellcolor{figamberlt}\textbf{65.0} \\
& Dict.\ $\times$5, CISA $\times$3, replay 2.5 & 405 & 28.1 & 4.31 & \textbf{99.5} & 18.5 & 1.91 & 31.7 \\
\bottomrule
\end{tabularx}
\caption{\textbf{Stage-2 arms} (new in v2). ``Facts'' is key-fact recall (\%); judge scores are 0--10. Values are means of two runs that differ only in dictionary row order; the baseline and the CISA-only arm are single runs. Replay ratio 1.2278 unless stated. ``CISA techn.'' is the share of seen-CTI technique questions answered with the advisory's techniques. Bold and underlined values mark the best and second-best stage-2 arm per column; the highlighted arm was selected for further work. Held-out T\# accuracy was 0\% for every run.}
\label{tab:stage2}
\end{table}

\begin{figure}[t]
\centering
\begin{subfigure}[b]{0.49\textwidth}\centering
\includegraphics[width=\linewidth]{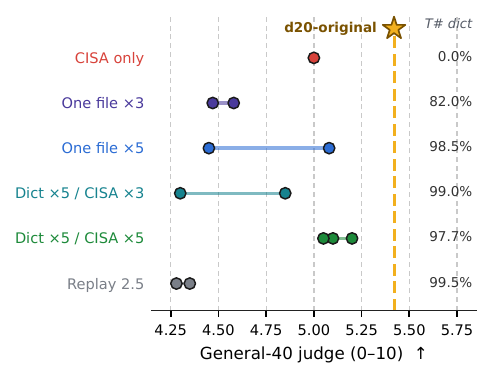}
\caption{General ability per arm}
\end{subfigure}\hfill
\begin{subfigure}[b]{0.49\textwidth}\centering
\includegraphics[width=\linewidth]{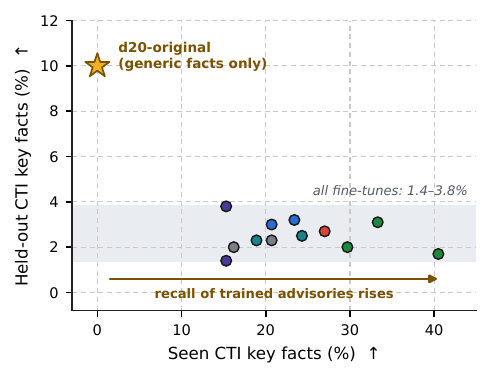}
\caption{Seen vs.\ held-out CTI}
\end{subfigure}
\caption{\textbf{Stage-2 fine-tuning learns what it is shown and little else.} One dot per run; colors mark the arm, as labeled in (a). \textbf{(a)} Every fine-tune scores below the baseline (star, dashed line) on held-out general prompts, and runs that differ only in row order spread widely; the right column gives each arm's mean in-dictionary T\# accuracy (three runs for ``Dict $\times$5 / CISA $\times$5''). \textbf{(b)} Recall of the trained advisories rises to 40\% of key facts, while knowledge of held-out groups, software, campaigns, and vulnerabilities stays at 1.4--3.8\%; the baseline's 10\% consists only of generic facts.}
\label{fig:stage2}
\end{figure}

\textbf{Results.} Table~\ref{tab:stage2} summarizes the arms, Figure~\ref{fig:stage2} shows every run, and Figure~\ref{fig:samples} gives verbatim answers. Five findings stand out.
\begin{enumerate}[topsep=2pt, itemsep=2pt]
  \item \emph{In-dictionary lookup is learnable in minutes.} Three epochs of the dictionary reached 79--85\% ID$\to$name accuracy (mean 82\%) in about 15 minutes of training; arms with five dictionary epochs averaged 97--99.5\%, stable across row orders. The remaining errors are confusions with nearby entries, mostly a sibling sub-technique under the correct parent or an ID one digit away; in the first run, only one of 15 misses invented a name.
  \item \emph{Held-out IDs are never named correctly.} Every run scored 0 of 50 on held-out IDs. The answers are names of real techniques, not invented ones: for the 36 held-out sub-techniques, the selected run (below) gave the correct parent ID in all 36 answers and a sibling's name in 16. Because the parent ID is a prefix of the asked ID and the dictionary teaches parent--child pairs explicitly, this shows confusion within a family rather than evidence that the hierarchy was learned. Placing the correct ATT\&CK entry in the prompt did not help: 78--84\% of answers collapsed into the CISA-advisory answer template, filled largely with technique IDs that are frequent in the CISA training answers, and at most 2 of 50 answers used the entry correctly.
  \item \emph{Seen content improves; unseen knowledge does not.} Recall of the training advisories rose from 0\% to as much as 35.1\% of key facts (65\% of technique questions answered with the advisory's techniques). Held-out CTI stayed at 1.8--3.1\% key facts for every arm, almost all of it generic facts such as ``malware'' or ``Windows'' (specific facts $\le$1.1\%); the baseline's 10\% consisted entirely of such generic facts. Stage-2 SFT on this data teaches the content it is given and nothing beyond it.
  \item \emph{Every fine-tune costs some general ability.} The general-40 judge score fell from 5.42 to 4.31--5.15 across the arm means, and validation loss rose by 0.019--0.025 bits per byte. Raising the replay ratio to 2.5 did not protect general ability: it nearly doubled the step count and gave the worst general-40 and validation loss.
  \item \emph{Training order is a large source of variance.} The ``dictionary $\times$5, CISA $\times$5'' arm and the one-file $\times$5 arm contain exactly the same rows and epochs and differ only in how the fixed shuffle ordered them, yet they differ by 13 points on seen CTI and 6.7 points on general-40 key facts. Two runs of the same arm differed by up to 10.8 points. Differences of about 10 points between two-run arms are therefore within order noise, while validation loss is stable to about $\pm$0.00005 and cannot rank these runs.
\end{enumerate}

\begin{figure}[p]
\centering
\begin{samplebox}{ATT\&CK lookup, in dictionary}{Exact-name hit: \textbf{Baseline} {\color{figbad}$\times$} \textcolor{figgray}{\,|\,} \textbf{Stage 2} {\color{figok}\checkmark}}
\slab{Prompt}Which MITRE ATT\&CK technique has the ID T1558.003?\par
\slab{Reference}Kerberoasting
\ssep
\slab{Baseline}The MITRE ATT\&CK technique has the ID T1558.003. This is a 1558 TT\&CK technique, which is a 1558 TT\&CK technique. The technique is \tagbad{token cap}\par
\slab{Stage 2}T1558.003 is Kerberoasting, a sub-technique of T1558 (Steal or Forge Kerberos Tickets). \tagok{correct}
\end{samplebox}
\begin{samplebox}{ATT\&CK lookup, held out}{Exact-name hit: \textbf{0} of 3 stage-2 answers}
\slab{Prompt}Which MITRE ATT\&CK technique has the ID T1001.001? \textcolor{figgray}{(also T1027.002; optionally after the correct entry)}\par
\slab{Reference}T1001.001 Junk Data; T1027.002 Software Packing
\ssep
\slab{Baseline}The MITRE ATT\&CK technique has the ID T1001.001. This is a 1001 T1001 T1001 T1001 T1001 \cut\par
\slab{Stage 2}T1001.001 is Protocol or Service Impersonation, a sub-technique of T1001 (Data Obfuscation). \tagmid{sibling}\par
\slab{Stage 2, entry given}Based on the CISA advisory, the following ATT\&CK techniques have been associated with this activity: T1001.001, T1001.002, T1001.003, T1001.00 \tagbad{template}\par
\slab{Stage 2, entry given}T1027.002 is ClickOnce, a sub-technique of T1027 (Obfuscated Files or Information). \tagmid{other technique}
\end{samplebox}
\begin{samplebox}{Seen CTI: training advisory}{Judge: \textbf{Baseline} 2 \textcolor{figgray}{\,|\,} \textbf{Stage 2} 7}
\slab{Prompt}Name three MITRE ATT\&CK technique IDs that CISA lists for the Rhysida Ransomware activity.\par
\slab{Reference}T1003.003, T1016, T1018, T1021.001, T1021.004, T1033, \ldots{} (24 IDs)
\ssep
\slab{Baseline}Here are three MITRE ATT\&CK technique IDs that CISA lists for the Rhysida Ransomware activity are: 1. CISA: CISA lists for the Rhysida Ransomware activity are as follows: - CISA: CISA lists for the Rhysida Ransomware activity are as follows: \cut\par
\slab{Stage 2}CISA lists for the Rhysida Ransomware activity are: T1003, T1003.001, T1003.003, T1003.005, T1003.007, T1003.008, T1003.009, T1003.0010, T1003.0011, T1010.002, T1010.003, T1012, T1012.002, T1018, T1021.001, T1021.004, \cut{} \tagok{3 of 3 key facts}
\end{samplebox}
\begin{samplebox}{Held-out CTI: threat group and vulnerability}{Key facts: \textbf{0\%} for both models}
\slab{Prompt}Who is the threat group BRONZE BUTLER?\par
\slab{Reference}aliases REDBALDKNIGHT, Tick; attributed to China; uses ABK, Avenger, BBK, Daserf, Mimikatz
\ssep
\slab{Baseline}The threat group BRONZE (Boside and Prego) is a group of individuals who have been affected by the threat of BRONZE (Boside and Prego) for a variety of reasons. \cut\par
\slab{Stage 2}The threat group BRONZE (Browser Group of Service) is a group of individuals and organizations that have been known to target and control a wide range of businesses and individuals, including those in the food and beverage industry, healthcare, manufacturing, and transportation. \cut
\ssep
\slab{Prompt}What is the vulnerability CVE-2020-16017?\par
\slab{Reference}Google Chrome; use-after-free; possible sandbox escape
\ssep
\slab{Baseline}The vulnerability CVE-2020-16017 is a complex and multifaceted issue that has been debated by scholars and researchers for centuries. \cut\par
\slab{Stage 2}CVE-2020-16017 is a vulnerability affecting the remote code execution environment of VMware VMware VMware VMware VMware \cut
\end{samplebox}
\begin{samplebox}{General-40: science}{Judge: \textbf{Baseline} 8 \textcolor{figgray}{\,|\,} \textbf{Stage 2} 8}
\slab{Prompt}Why does the sky look blue during the day?
\ssep
\slab{Baseline}The sky appears blue during the day because of a phenomenon known as Rayleigh scattering. When sunlight enters the Earth's atmosphere, it encounters tiny particles in the atmosphere, such as dust, pollen, and pollutants. \cut\par
\slab{Stage 2}The sky appears blue during the day because of a phenomenon known as Rayleigh scattering. When sunlight enters the Earth's atmosphere, it encounters tiny particles of blue and violet that scatter the shorter blue and violet light. \cut
\end{samplebox}
\caption{\textbf{Verbatim greedy answers} from the baseline (d20-original) and the selected stage-2 run. Items follow a rule fixed before the answers were read (first item of each test set, plus T1558.003 and T1001.001). \cut{} marks omitted text, mostly repetition loops. ``Entry given'': the correct ATT\&CK entry preceded the question. The v1 CTI checkpoint was not evaluated on these test sets.}
\label{fig:samples}
\end{figure}

A third run of the ``dictionary $\times$5, CISA $\times$5'' arm gave a three-run mean of 97.7\% T\# accuracy, 5.12 general-40 judge score, and 34.5\% seen-CTI key facts. A selection rule fixed before that run (T\# accuracy $\ge$95\%, then the highest general-40 judge score, provided seen CTI is not the lowest of the three) selects the first run, by a margin of 0.10--0.15 judge points against a judge-only noise of about 0.07 per item. The overall conclusion is that a 0.9-billion-parameter model trained this way memorizes the dictionary and the advisories it sees, but does not generalize to unseen identifiers or entities. Follow-on work on extracting techniques from unseen CISA advisory text will be reported separately.

\section{Embedding-Space Inspection}
\label{sec:embedding}

A companion analysis projected the NanoChat token-embedding matrix with t-SNE. The presentation records a 128-dimensional embedding space and a vocabulary of approximately 32{,}000 tokens. Qualitative clusters included natural-language vocabulary, educational terms, subword fragments, reasoning/numerical contexts, and technical/formal vocabulary. This visualization is a useful diagnostic that the trained model organized tokens non-randomly, but it is not a quantitative evaluation of language-model quality. The slide deck labels the illustrated model as approximately 31 million parameters; this refers to a different, smaller NanoChat configuration than the depth-20 production run, whose parameter count is 896{,}533{,}746 (Table~\ref{tab:params}).

\section{Educational Use of the Cluster}
\label{sec:education}

Beyond the research workloads above, the same two-node cluster supports instructional use in two Grambling State University courses, all reached over the same Tailscale tailnet described in Section~\ref{sec:architecture}. We report this as an operational deployment note rather than a controlled pedagogical study: no pre/post assessment or outcome comparison has been conducted, and the discussion below is qualitative.

\subsection{CS 426: Artificial Intelligence}
In CS~426 (Artificial Intelligence), a 400-level undergraduate course, students are given hands-on exposure to the same NanoChat pipeline documented in Sections~2 and~4 -- tokenization, base pretraining, supervised fine-tuning, and interactive inference. Rather than treating transformer training as a black box, students inspect intermediate artifacts (tokenizer vocabulary, training loss curves, checkpoint files) produced on the cluster, connecting lecture material on embeddings and self-attention to the embedding-space inspection in Section~\ref{sec:embedding} and to infrastructure they can run themselves rather than only read about, over their own Tailscale-connected devices.

\subsection{CBS 255: CompTIA Security+ POGIL Activities}
CBS~255 (Spring 2026) uses Process Oriented Guided Inquiry Learning (POGIL) activities~\cite{pogil} converted from CompTIA Security+ (SY0-701) lecture material across ten lessons, each with model-based directed/convergent/divergent questions and Performance-Based Question (PBQ) scenarios (drag-and-drop and matching exercises mirroring the exam's PBQ format). These human-authored materials are served to students through a purpose-built Flask web application (126 model-based question/answer pairs plus 88 PBQ scenario questions across the ten lessons, 214 total) that also exposes an AI practice-quiz mode: for a selected lesson and difficulty level, an Ollama-hosted \texttt{llama3:latest} model~\cite{ollama} generates a fresh, previously unseen four-option multiple-choice question on that lesson's topic, distinct from the fixed POGIL questions, with immediate feedback and an explanation. Difficulty levels follow Bloom's taxonomy~\cite{bloom} -- Recall, Apply, and Analyze -- and are mapped to workshop timing: Recall during the POGIL activity itself, Apply immediately afterward, and Analyze as exam-prep review, since SY0-701 items are predominantly written at the Apply/Analyze level. The web application (hosted on a campus server) and the Ollama instance are on separate machines: Ollama runs on \texttt{spark-9fcd} -- rank~0 of the same two-node cluster described in Section~\ref{sec:architecture} -- and is reached over the existing Tailscale management network rather than a public API, so certification-practice traffic never leaves university-controlled infrastructure. Material is distributed to students via Google Classroom as a self-check reference (answer keys) and a graded assignment (AI practice quiz, submitted by screenshot). As with CS~426, this is reported as an operational deployment note; a systematic comparison of learning outcomes with and without the AI practice-quiz mode has not been conducted, and AI-generated questions were manually spot-checked rather than exhaustively validated against CompTIA's exam blueprint.

\subsection{Discussion}
Both instructional uses share infrastructure with the research workloads in Sections~5 and~6 rather than requiring separate hardware. Notably, the CBS~255 Ollama endpoint runs on \texttt{spark-9fcd}, the same rank-0 node used for the depth-20 pretraining run in Section~5 -- the identical Tailscale-managed network (Section~\ref{sec:architecture}) that administers distributed training also carries classroom inference traffic. This is consistent with the motivation in Section~1: modest local infrastructure can support multiple, otherwise-siloed activities -- distributed pretraining research, downstream domain fine-tuning, and certification-aligned instruction -- on the same two nodes, which is a practical argument for this class of hardware, paired with a mesh-VPN remote-access layer, in resource-constrained departments even before any scaling-efficiency claim is established.

\section{Operational Lessons}

The proof-of-concept produced several practical lessons:
\begin{itemize}[topsep=2pt, itemsep=2pt]
  \item \textbf{Separate management and data planes.} Tailscale simplified remote access, but NCCL was forced onto the dedicated 200~Gb/s interface.
  \item \textbf{Persist network configuration.} Manual \texttt{ip addr add} commands do not survive reboot; the two systems required persistent NetworkManager/netplan configuration.
  \item \textbf{Use host networking and sufficient memlock.} The container must see the physical interface, and NCCL requires appropriate memory-lock and stack limits.
  \item \textbf{Treat compile/startup time separately.} GB10 kernel compilation/autotuning can exceed default watchdog expectations and should not be mixed with steady-state throughput.
  \item \textbf{Checkpoint early and verify.} A default ``save only at the end'' policy is unacceptable for multi-day experiments. The production run itself ran with \texttt{save\_every=-1} and kept only its final checkpoint; verify the recorded configuration, not the intended one.
  \item \textbf{Rank silence is not rank failure.} Only rank~0 logs progress by convention; rank~1 should be checked through GPU utilization and process memory.
  \item \textbf{Record exact software state.} The final artifact must include the NanoChat commit hash, container digest, local patches, complete launch command, and raw rank~0 log.
  \item \textbf{Count gradient accumulation.} Tokens per step must be computed from the optimizer batch, not the micro-batch; v1's eightfold undercount came from this.
  \item \textbf{Audit the write-up against the artifacts.} Checkpoint metadata, training reports, shell histories, and saved evaluation files exposed every error corrected in v2. Comparing a draft against them before release is cheap.
\end{itemize}

\section{Threats to Validity and Limitations}
\label{sec:limitations}

This report has several limitations. First, it is a two-node case study on one hardware and software configuration. Second, no single-node depth-20 run was logged, so no scaling claim is made, and the earlier single-node and d12 figures in Table~\ref{tab:phases} remain unverified. Third, the raw rank-0 training log was not retained; v2's corrections rest on checkpoint metadata, nanochat's training reports, and shell histories. Fourth, Python packages were installed ad hoc in each container, and although the histories indicate the NGC PyTorch build was used, they cannot prove which interpreter launched each run (Section~\ref{sec:repro}). Fifth, the v1 CTI evaluation contains only 17 prompts, sampled answers, and one LLM judge, with no human ratings, independent judges, confidence intervals, or inter-rater agreement. Sixth, the stage-2 study uses one 0.9-billion-parameter model, two or three runs per arm, and a single judge model; the base model's pretraining data was not audited for exposure to the held-out IDs. Seventh, the educational-use observations in Section~\ref{sec:education} are qualitative operational notes, not a controlled pedagogical evaluation. Finally, the photograph in Figure~\ref{fig:lab} documents the physical deployment but does not verify traffic routing or performance; those claims depend on the runbook and logs described in Section~4.

\section{Reproducibility Package}
\label{sec:repro}

The current package contains the following artifacts:
\begin{itemize}[topsep=2pt, itemsep=1pt]
  \item dual-node DGX Spark setup runbook, including Tailscale ACL configuration;
  \item laboratory setup photograph;
  \item CTI train and validation JSONL files;
  \item CISA-to-SFT conversion script;
  \item held-out validation set;
  \item automated response-generation, merge, and LLM-judge scripts;
  \item CTI evaluation summary; and
  \item NanoChat embedding visualization.
\end{itemize}

\textbf{Software state (new in v2).} Pretraining and both v1 SFT runs used NanoChat commit \texttt{dc54a1a} with two local patches: the step-zero evaluation guard (Section~4.3) and a guard on the last step of distributed SFT in \texttt{chat\_sft.py}, which v1 did not mention. The container was \nolinkurl{nvcr.io/nvidia/pytorch@sha256:42263b2424fc237b34c4fc4a91c30d603c57eed36e37d31ff6d9a4f1f801edee} (torch \texttt{2.9.0a0+145a3a7bda.nv25.10}, Python 3.12.3). Supporting packages such as \texttt{datasets}, \texttt{tiktoken}, and \texttt{tokenizers} were pip-installed in each container, and their versions varied between containers. The stage-2 study used ATT\&CK Enterprise v20.0 from the \texttt{attack-stix-data} repository at commit \texttt{20664a889568} (file sha256 \texttt{81113ba9\ldots}).

A more complete public release would additionally include \texttt{launch\_rank0.sh}, \texttt{launch\_rank1.sh}, the local patches, system-information snapshots, and a machine-readable results table -- a sanitized subset of which is included in the public repository described under Code Availability below. The private network addresses reported in this paper refer only to the non-routable \texttt{192.168.100.0/24} training subnet and carry no external exposure. The judge endpoint, hard-coded in v1's \texttt{judge\_eval.py}, is a configuration setting in the course-kit version used for the stage-2 study.

\section{Conclusion}

This work demonstrates that two compact DGX Spark systems can be configured as a functional distributed LLM training cluster using PyTorch \texttt{torchrun}, DDP, NCCL, and a dedicated 200~Gb/s direct link, with all remote administration carried over a Tailscale mesh VPN. The 896.5-million-parameter depth-20 NanoChat run processed 1{,}048{,}576 tokens per optimizer step and averaged approximately 16{,}540 aggregate tokens/s of training time over a four-day, 5.22-billion-token pretraining run. The deployment artifacts capture the engineering details that made the run possible, including persistent network setup, interface binding, container configuration, timeout handling, checkpointing, and rank monitoring. A complete CISA CTI fine-tuning and evaluation workflow demonstrates one downstream use of the trained checkpoint; the stage-2 study added in v2 shows that the model readily memorizes an ATT\&CK dictionary and the advisories it is trained on but does not generalize to unseen identifiers or entities, and concurrent use by a 400-level AI course and a Security+ POGIL/PBQ query engine demonstrates that the same infrastructure -- and the same remote-access testbed -- serves teaching as well as research.

The study establishes feasibility and reproducibility rather than a definitive scaling-efficiency result: as discussed in Section~\ref{sec:training}, a short, matched one-node/two-node benchmark would be needed to convert this proof of concept into a controlled performance study, and we have deliberately avoided claiming that result here. In its current form, the report is offered as a reproducibility and deployment reference for small laboratories building local, privacy-preserving distributed LLM infrastructure -- accessible remotely over a lightweight mesh VPN -- that can be shared between research and instruction.

\subsection*{Artifact Availability}
The runbook, CISA-to-SFT conversion script, held-out evaluation set, and evaluation scripts described in this report are maintained by the corresponding author; a sanitized subset is publicly available (see Code Availability). No cloud service, external API, or third-party dataset requiring separate licensing is needed to reproduce the deployment procedure in Section~4; the CISA advisory corpus underlying Section~6 is publicly available from CISA~\cite{cisa}. MITRE ATT\&CK data are \copyright\ The MITRE Corporation and are used under MITRE's terms of use~\cite{mitre}.

\subsection*{Acknowledgments}
This material is based upon work supported by the National Science Foundation under Grant No.\ 2101181. Any opinions, findings, and conclusions or recommendations expressed in this material are those of the author(s) and do not necessarily reflect the views of the National Science Foundation. This work was conducted under a collaborative NSF award between Grambling State University (PI: Vasanth Iyer) and the University of North Texas (Co-PIs: Saraju P.\ Mohanty and Elias Kougianos). We thank Harvey Farley (Department of Computer Science \& Digital Technologies, Grambling State University) for his support in establishing the AIoT Lab infrastructure on which these experiments were conducted.

\subsection*{Code Availability}
The deployment runbook, network-architecture documentation, and troubleshooting notes described in Sections~3--4 and~9 are maintained as a public, sanitized reference repository at \url{https://github.com/viyer-research/dual-dgx-spark-tailscale-testbed}, released under the MIT License.

\end{document}